\documentclass[reprint,onecolumn,notitlepage,amsmath,amssymb,aps]{revtex4-1}

\usepackage{graphicx}
\usepackage{dcolumn}
\usepackage{bm}

\usepackage{amsmath} 
\usepackage{amsthm}  

\begin{document}

\title{Memory-Retaining RG Fixed Points in Nonlinear PDEs}

\author{Ko Okumura\\ 
\small E-mail: okumura.ko@ocha.ac.jp\\ 
\small Department of Physics and Soft Matter Center, Ochanomizu University, 2-1-1 Ohtsuka, Bunkyo-ku, Tokyo 112-8610, Japan}

\date{\today}

\begin{abstract}

Self-similar solutions of partial differential equations (PDEs) are often
interpreted as fixed points of renormalization-group (RG) transformations.
Let $\xi $ denote a similarity variable and $\eta $ a dimensionless physical
scale. In conventional RG descriptions, asymptotic coarse graining leads to
memory loss, whereby the physical scale $\eta $ disappears from the
fixed-point solution. In this work, we develop an RG framework in which
fixed points can take the more general form $\eta ^{\alpha }\Gamma (\xi
/\eta ^{\beta })$, where the surviving scale $\eta $ remains embedded in the
asymptotic solution.

While the case $\beta =0$ with $\alpha \neq 0$, corresponding to anomalous
scaling through amplitude renormalization $\eta ^{\alpha }$, has long been
understood within RG theory, self-similar solutions with $\beta \neq 0$ are
well known in studies of self-similarity of the second kind, but have lacked
a general RG formulation. We develop a nonperturbative RG framework for
nonlinear PDEs that extends RG fixed-point theory to self-similar solutions
with $\beta \neq 0$ by treating physical scales as dynamical variables of
the RG flow.

The resulting framework identifies which scales remain asymptotically
relevant and leads naturally to three classes of RG fixed points:
memory-loss, memory-retaining I, and memory-retaining II. In the first
class, scale information disappears from the asymptotic solution. In the
second, it survives only through multiplicative amplitude factors, such as $%
\eta ^{\alpha }$. In the third, corresponding to $\beta \neq 0$, it remains
embedded in the similarity profile $\Gamma $ itself.

The framework is illustrated using conservative diffusion, the Barenblatt
diffusion equation, a modified density-dependent diffusion model, and a
fatigue-crack-growth problem involving three physical length scales. These
examples demonstrate how RG flows determine both the surviving scales and
the structure of the asymptotic self-similar solution, providing a unified
RG framework for memory retention and, in particular, for self-similar
solutions of the second kind with $\beta \neq 0$.

\end{abstract}

\maketitle

\section{Introduction}

\label{sec:intro}

Self-similar solutions are ubiquitous in nature and constitute one of the
most important manifestations of universality \cite{Barenblatt, Goldenfeld}.
They arise in a wide variety of nonlinear PDEs, including porous-medium,
thin-film, Keller--Segel, and aggregation equations, among many others.
Barenblatt classified self-similar solutions into two categories, the first
and second kinds, depending on whether the scaling exponents can be
determined by dimensional analysis. Motivated by Barenblatt's work,
Goldenfeld and co-workers developed a so-called field theoretic RG theory
for nonlinear PDEs and showed that self-similar solutions of the second kind
can be understood in terms of anomalous dimensions \cite%
{goldenfeld1989intermediate, Goldenfeld}, establishing a connection with 
\textit{renormalization} in quantum field theory (QFT) \cite%
{ryder1996quantum}. The resulting self-similar solutions are characterized
by anomalous exponents, while the asymptotic profile remains a function of a
single similarity variable, corresponding to the $\beta =0$ case.

More recently, a unified RG framework has been proposed for nonlinear PDEs 
\cite{Okumura2025RG, okumura2026combined, Okumura2026nonlinear}. The
framework is based on repeated scale transformations and associated RG flow
equations \cite{bricmont1994renormalization}. Within this formulation,
self-similar solutions appear as RG fixed points, while universality classes
emerge through the elimination of irrelevant structures, thereby extending
the RG logic underlying critical phenomena to nonlinear PDEs.

The framework unifies the field-theoretic RG theory with the BKL-type RG
theory, introduced by Bricmont, Kupiainen, and Lin \cite%
{bricmont1994renormalization}, and the dynamical-system description (DSD),
pioneered by Giga and Kohn \cite{giga1985asymptotically} and developed by
others \cite{eggers2015singularities}. More broadly, it extends RG
principles underlying critical phenomena \cite{Cardy, Goldenfeld} to
nonlinear PDEs, allowing asymptotic scaling and universality classes to be
predicted directly from the governing equations.

Unlike Hamiltonian-based RG, where coarse-graining eliminates microscopic
degrees of freedom and fixed points reside in coupling-constant space,
renormalization in nonlinear PDEs for a dynamical field $H(T,\mathbf{X})$
acts directly on the field $H(T,\mathbf{X})$ itself. Here, we formulate the
RG transformation on an extended state vector $\mathbf{X}$ that includes not
only spatial coordinates but also all competing physical length scales, such
as those characterizing initial conditions, material properties, or system
geometry. Using the scaling dimensions $A$ and $B$ of $H$ and $T$,
respectively (Sec \ref{sec:rg}), the unified RG framework predicts
self-similar RG fixed-point solutions of the form $H^{\ast }(T;\mathbf{X}%
)=T^{-A/B}$ $f(\xi ,\boldsymbol{\eta })$ with $\xi =X/T^{1/B}$ and $\eta
_{i}=X_{i}/T^{1/B}$. Here $\xi $ is the rescaled spatial similarity
coordinate and $\boldsymbol{\eta }=(\eta _{0},\eta _{1},\dots )$ denotes the
rescaled physical length scales, enabling self-similar solutions beyond the
conventional $\beta =0$ case.

In self-similarity of the first kind, all auxiliary scales $\boldsymbol{\eta 
}$ drop out of $f(\xi ,\boldsymbol{\eta })$ asymptotically, leaving a
single-variable fixed-point function $f=\Gamma (\xi )$ that loses memory of
the initial condition. In self-similarity of the second kind, by contrast,
some components of $\boldsymbol{\eta }$ survive asymptotically. In
conventional field-theoretic RG approaches \cite%
{Goldenfeld,goldenfeld1989intermediate,chen1994renormalization}, the
fixed-point profile remains a function of the single similarity variable $%
\xi $, even when anomalous scaling exponents are present. In contrast, the
extended-state-space RG developed here allows fixed points to retain
explicit dependence on asymptotically relevant scales, leading to the
unified structure $f(\xi ,\boldsymbol{\eta })=\eta ^{\alpha }\Gamma (\xi
/\eta ^{\beta })$. Here, $\beta =0$ corresponds to amplitude-only retention,
whereas $\beta \neq 0$ describes profile-reshaping retention in which the
surviving scale modifies the similarity argument itself.

Within this representation of the fixed-point function, two fundamental
questions arise. First, in self-similarity of the second kind, no general
principle has been available for determining which physical scales survive
asymptotically and how the surviving scales enter the fixed-point solution.
In particular, it has remained unclear whether scale information appears
only through multiplicative amplitudes or also modifies the similarity
profile itself. As a result, conventional analyses typically rely on
case-specific similarity ansatz \cite{Barenblatt,barenblatt2003scaling}.

Second, from the viewpoint of RG theory, fixed points in which surviving
scales remain explicit arguments of the asymptotic profile have not been
formulated within a general nonperturbative RG framework. Consequently,
self-similar solutions corresponding to the $\beta \neq 0$ class have lacked
a unified RG description.

In this paper, we resolve both questions within the unified predictive PDE
RG framework \cite{Okumura2025RG, okumura2026combined, Okumura2026nonlinear}%
. By nondimensionalizing the governing equations using an arbitrary external
reference length rather than an intrinsic problem scale, all physical scales
are retained explicitly in the RG state space and treated on an equal
footing. The resulting stationary RG equation possesses two invariances:
unit-choice invariance with respect to the arbitrary reference length and a $%
\lambda $-rescaling invariance. Together, these invariances determine which
physical scales remain asymptotically relevant and which are eliminated
under coarse graining.

This scale-selection mechanism fixes the scaling dimensions $A$ and $B$ (or
their invariant ratio), determines the anomalous exponents $(\alpha ,\beta )$%
, and predicts whether surviving scales appear only through multiplicative
amplitudes ($\beta =0$) or remain embedded in the similarity profile ($\beta
\neq 0$). All results are obtained nonperturbatively and without requiring
an a priori similarity ansatz.

To illustrate the resulting fixed-point structures and associated
scale-selection mechanism across both two-scale solvable PDEs and multiscale
problems with competing initial, material and geometric scales, we apply the
framework to a hierarchy of three systems: (i) conservative linear and
density-dependent diffusion (DDD) models \cite{Barenblatt,
murray2002mathematical1}, where it combines with mass conservation to
eliminate the initial length scale and select the classic \textit{memory-loss%
} fixed point; (ii) a nonconservative modified DDD model \cite%
{barenblatt2003scaling}, where it preserves the initial scale to yield an
exact \textit{memory-retaining II} fixed point with logarithmic cutoff
signatures, while recovering Barenblatt's nonlinear diffusion equation \cite%
{Barenblatt} in the $\beta =0$ limit as a \textit{memory-retaining I} fixed
point; and (iii) Barenblatt's three-scale fatigue-crack growth problem \cite%
{Barenblatt, barenblatt2014flow}, where the initial crack length $l_{0}$,
microscopic fracture scale $l_{c}$, and specimen size $h$ compete, and the
same mechanism eliminates $l_{0}$ while combining $l_{c}$ and $h$ into the
spatial-free counterpart of a \textit{memory-retaining II} fixed point that
recovers Paris's scaling law, even though the governing equation is not
completely specified.

The remainder of this paper is organized as follows. Section \ref{sec:rg}
summarizes the unified RG framework. Sections \ref{sec:ddd} and \ref%
{sec:mddd} analyze the conservative and nonconservative DDD models,
respectively, and Section \ref{sec:fatigue} presents the three-scale
extension to fatigue-crack growth. Section \ref{sec:discussion} discusses
the resulting fixed-point structures, classification, and broader
implications, and Section \ref{sec:conclusion} concludes. Technical
derivations are collected in Appendix A.

\section{\noindent Unified RG framework}

\label{sec:rg}

In previous studies \cite{Okumura2025RG, okumura2026combined,
Okumura2026nonlinear}, the unified RG framework has been developed for a set
of PDEs for the vector field $\mathbf{H}(T;\mathbf{X})$. In this work, we
consider the case in which the vector field only has one component $H(T;%
\mathbf{X})$ with a single spatial variable $X$ governed by%
\begin{equation}
\partial _{T}H(T;\mathbf{X})=F(H,\partial _{X}H,\partial _{X}^{2}H,\cdots
)+G(H,\partial _{X}H,\cdots ),  \label{eq13}
\end{equation}%
for which, we introduce a BKL-type scale transformation for a scale factor $%
L $ and the exponent $B$, which are both positive:

\begin{eqnarray}
T^{\prime } &=&T/L^{B},\text{ }\mathbf{X}^{\prime }=\mathbf{X}/L
\label{eq14} \\
H^{\prime }(T^{\prime };\mathbf{X}^{\prime }) &=&L^{A}H(T,\mathbf{X})\text{ }%
\equiv H^{L}(T^{\prime },\mathbf{X}^{\prime })  \label{eq15} \\
&\Leftrightarrow &\text{ }H^{L}(T,\mathbf{X})=L^{A}H(L^{B}T,L\mathbf{X}).
\label{eq16}
\end{eqnarray}%
\newline
The present formulation emphasizes that all relevant length scales may be
regarded as components of the vector $\mathbf{X}=(X,X_{0},X_{1},\ldots )$.
The term in $F$ represents \textit{scale-invariant} terms, whereas $G$
non-scale-invariant terms, with the latter \textit{irrelevant} in the sense
that they are eliminated as the RG transformation defined in Eq \ref{eq17}
is repeatedly applied: Eq \ref{eq13} with zero $G$ defines a \textit{%
universality class} and Eq \ref{eq13} with nonzero irrelevant $G$ belongs to
the universality class defined by Eq \ref{eq13} with $G=0$. The scale factor 
$L$ is in the range $0<L<1$ for singular dynamics (\textit{short-time
asymptotics}) where \textit{small scale physics} is important (Case I),
whereas it satisfies $L>1$ for \textit{long-time asymptotics} where \textit{%
large scale physics} is important (Case II). The terms in $F$ and $G$ are
regarded as \textit{scale-invariant}, \textit{relevant}, and \textit{%
irrelevant} depending on whether the exponent $M$ defined in Eq \ref{M1} is
zero, negative (positive), and positive (negative), respectively, for Case I
(Case II). Furthermore, as detailed in Appendix \ref{A-U}, the framework
shows that \textit{self-similar solutions emerge as RG FPs if }$f(\mathbf{X}%
) $ \textit{exists}: 
\begin{equation}
H^{\ast }(T;\mathbf{X})=T^{-A/B}f(\bm{\xi })\text{ with }\bm{\xi }=\mathbf{X}%
/T^{1/B}  \label{e20}
\end{equation}%
where $f(\bm{\xi })$ is the stationary solution to the RG flow equation
given in Eq \ref{eq21}: 
\begin{equation}
0=A\widetilde{f}(\mathbf{X})+BF(\widetilde{f},\partial _{X}\widetilde{f}%
,\partial _{X}^{2}\widetilde{f},\cdots )+\sum_{i}X_{i}\frac{\partial 
\widetilde{f}(\mathbf{X})}{\partial X_{i}}.  \label{eq21A}
\end{equation}

\section{\noindent DDD model}

\label{sec:ddd}

We consider below a long-time asymptotics of biological DDD model \cite%
{murray2002mathematical1}, which is also called the PMD model \cite%
{Barenblatt}:%
\begin{equation}
\partial _{t}h\mathbf{(}t,x)=2\kappa \partial _{x}(h\partial _{x}h)
\label{eq24}
\end{equation}%
under the mass conservation $\int_{-\infty }^{\infty }h(t,x)dx=m_{0}$ with a
very intense initial distribution of the form $h\mathbf{(}0,x)=\frac{m_{0}}{l%
}f_{0}(\frac{x}{l})$. The function $f_{0}(X)$ is an even function of $X$ and
decays quickly with $X$ with a characteristic length $l$, as typically
expressed by the Gaussian distribution. Introducing a unit of length $l_{0}$
and dimensionless quantities $H=\frac{h}{h_{0}}$ with $h_{0}=\frac{m_{0}}{%
l_{0}}$, $X=\frac{x}{l_{0}}$, and $T=t/t_{0}$ with $1/t_{0}=\kappa
m_{0}/l_{0}^{3}$, we obtain dimensionless forms:

\begin{equation}
\partial _{T}H\mathbf{(}T,X)=2\partial _{X}(H\partial _{X}H)  \label{DDD}
\end{equation}%
and $\int_{-\infty }^{\infty }H(T,X)dX=1$.

We apply the above-introduced unified RG framework for Case II ($L>1$) to Eq %
\ref{DDD} by identifying this equation with Eq \ref{eq13}. In principle, $%
\mathbf{X}$ in the unified framework includes all the relevant length scales
in the problems. Intuitively, however, in the present case, at long times
where large scale physics is essential, we naturally expect that the
solution may \textit{lose the memory} of the initial length scale $l$ (or
its dimensionless counterpart $X_{0}=l/l_{0}$. We therefore first neglect
the dependence of $H$ on $X_{0}$, anticipating complete \textit{scale
separation} in the long-time limit, and consider $H(T,X)$ instead of $%
H(T;X,X_{0})$.

Importantly, as shown in Appendix \ref{A-DD}, this elimination of $X_{0}$
does not need to be assumed a priori: even if we retain $X_{0}$ from the
outset by starting with $H(T;X,X_{0})$, the stationary RG flow equation
combined with mass conservation automatically forces the anomalous exponent
to vanish ($\beta =0$), confirming self-consistently that the initial length
scale drops out of the asymptotic fixed point, realizing a memory-loss fixed
point ($\alpha =\beta =0$).

The scale invariance of the left-hand side of Eq \ref{DDD} and the mass
conservation respectively give $B=A+2$ and $B=3$ (see Appendix \ref{A-DD}).
Thus we obtain the DDD scaling, $H^{\ast }(T,X)$ $=$ $T^{-1/3}f(X/T^{1/3})$,
from Eq \ref{e20}. The framework predicts that the\textit{\ universality
class} in the present case consists of PDEs of the form $\partial _{T}H%
\mathbf{(}T,X)=2\partial _{X}(H\partial _{X}H)+G$, where $G$ is given by a
linear combination of terms that can be expressed as $H^{n_{0}}(\partial
_{X}H)^{n_{1}}(\partial _{X}^{2}H)^{n_{2}}(\partial _{X}^{3}H)^{n_{3}}\cdots 
$ with a \textit{negative} $M$ in Eq \ref{M1} with $B=3$ and $A=1$. From the
RG flow equation in Eq \ref{eq21}, we can show the stationary solution $%
\widetilde{f}(X)=f(X)$ exists and is given by

\begin{eqnarray}
H^{\ast }(T,X) &=&T^{-1/3}f(X/T^{1/3})  \label{eq5} \\
&=&\frac{1}{12T^{1/3}}(X_{f}^{2}-(\frac{X}{T^{1/3}})^{2}),  \label{eq6}
\end{eqnarray}%
with $X_{f}=9^{1/3}$ for $\left\vert X\right\vert <X_{f}T^{1/3}$ and $%
H^{\ast }(T,X)=0$ otherwise (see Appendix \ref{A-DD}). In a dimensional
form, it is given as 
\begin{equation}
h^{\ast }(t,x)=\frac{3^{1/3}m_{0}^{2/3}}{4(\kappa t)^{1/3}}(1-\frac{x^{2}}{%
(9\kappa m_{0}t)^{2/3}})  \label{eq6a}
\end{equation}%
for $\left\vert x\right\vert <X_{f}(\kappa m_{0}t)^{1/3}$, which reproduces
a known result in a completely different manner, thereby providing a
nontrivial validation of the unified RG formulation. Notably, Eq \ref{eq6a}
depends on $m_{0}$ but is independent of $l$---reflecting the complete loss
of memory of the initial length scale at this memory-loss fixed point ($%
\alpha =\beta =0$), where only the conserved total mass $m_{0}$ governs the
asymptotic state. As we show next, once mass conservation is broken, this
memory loss fails and $l$ remains explicitly encoded in the asymptotic
solution.

\section{\noindent Modified DDD model}

\label{sec:mddd}

We consider the following modified DDD model

\begin{equation}
\partial _{T}H\mathbf{(}T,X)=2[\partial _{X}(H\partial _{X}H)-c(\partial
_{X}H)^{2}]  \label{eq2}
\end{equation}%
with $0<c<1$ and seek a long time asymptotics as before for the initial
distribution characterized by the length scale $X_{0}$. The additional
gradient-squared term is introduced as the simplest scale-invariant
nonconservative extension of the DDD equation. Besides its minimal RG
structure, this model is of particular interest because it corresponds to
one of the rare nonlinear eigenvalue problems discussed by Barenblatt \cite%
{barenblatt2003scaling} for which an exact analytical solution is known. It
therefore provides an ideal testing ground for examining the relation
between anomalous scaling and memory retention within the unified RG
framework.

In this modified model, as already mentioned, the mass is no longer
conserved: $\partial _{T}\int_{-\infty }^{\infty }H(T,X)dX=-2c\int_{-\infty
}^{\infty }(\partial _{X}H)^{2}$ $dX<0$. Thus, the exponent $A$ cannot be
determined by the dimensional analysis: this problem is the second kind of
self similarity in the terminology of Barenblatt. Furthermore, if we assume
the self-similar form $H(T,X)=T^{-A/B}f(X/T^{1/B})$ with $B=A+2$, we can
show $\int_{-\infty }^{\infty }H(T,X)dX$ is a constant, which contradicts
with non-conserved mass. This means that Eq \ref{eq2} does not admit a
self-similar solution of this form.

This contradiction suggests that the asymptotic solution cannot be described
solely in terms of the similarity variable $X/T^{1/B}$, although large-scale
physics should be important at long times. We therefore expect that the
initial length scale $X_{0}$ remains asymptotically relevant, and that 
\textit{memory of }$X_{0}$\textit{\ is anomalously retained} in this case:
the dependence on $X_{0}$ of the asymptotic solution $H^{\ast }(T,X,X_{0})$
cannot be neglected any more. In such a case, the unified RG framework tells
us that (1) the correct RG FP may be in the following form: 
\begin{equation}
H^{\ast }(T,X,X_{0})=T^{-A/B}f(\xi ,\eta ),  \label{eq3a}
\end{equation}%
where $\xi =X/T^{1/B}$, $\eta =X_{0}/T^{1/B}$, and $B=A+2$ (because the
extra term $\left( \partial _{X}H\right) ^{2}$ is also scale-invariant; see
Appendix \ref{A-MDD}), and that (2) the RG flow equation is given as $B\frac{%
d\widetilde{f}(\tau ,X,X_{0})}{d\tau }$ $=A\widetilde{f}$ $+2B$ $[\partial
_{X}(\widetilde{f}\partial _{X}\widetilde{f})-c(\partial _{X}\widetilde{f}%
)^{2}]$ $+X\frac{\partial \widetilde{f}}{\partial X}$ $+X_{0}\frac{\partial 
\widetilde{f}}{\partial X_{0}}$ and its stationary FP solution $\widetilde{f}%
=f$ satisfies 
\begin{equation}
Af(\xi ,\eta )+2B[\partial _{\xi }(f\partial _{\xi }f)-c(\partial _{\xi
}f)^{2}]+\xi \frac{\partial f}{\partial \xi }+\eta \frac{\partial f}{%
\partial \eta }=0  \label{eq11}
\end{equation}

Rather than imposing a complete space--time similarity ansatz on the
original PDE from the outset, we solve Eq \ref{eq11} by what we refer to as t%
\textit{he }$\lambda $\textit{-rescaling invariance method}, combined with
unit-choice invariance. While $H^{\ast }$ in Eq \ref{eq3a} is self-similar
in the extended $(\xi ,\eta )$ plane under simultaneous rescaling of $X$ and 
$X_{0}$, in a physical initial-value problem $X_{0}$ is held fixed as time
evolves, so $\eta =X_{0}/T^{1/B}$ varies with $T$. Consequently, as shown in
Appendix \ref{A-MDD}, for $H^{\ast }$ in Eq \ref{eq3a} to represent a
spatially self-similar profile at fixed $X_{0}$ under time evolution, $f(\xi
,\eta )$ must satisfy the scale-covariance identity ($\lambda >0$)%
\begin{equation}
f(\lambda ^{\beta }\xi ,\lambda \eta )=\lambda ^{\alpha }f(\xi ,\eta )
\label{eq3b}
\end{equation}%
with exponents $\beta $ and $\alpha $ to be determined. Setting $\lambda
=1/\eta $ in Eq \ref{eq3b} then directly yields, without assuming separation
of variables,

\begin{equation}
f(\xi ,\eta )=\eta ^{\alpha }\Gamma (\zeta )\text{ with }\zeta =\xi /\eta
^{\beta },  \label{eq23}
\end{equation}%
where $\Gamma (\zeta )=f(\zeta ,1)$. Furthermore, for $f(\xi ,\eta )$
satisfying Eq \ref{eq3b} to remain a solution of Eq \ref{eq11} under the $%
\lambda $-rescaling $(\xi ,\eta ,f)\rightarrow (\lambda ^{\beta }\xi
,\lambda \eta ,\lambda ^{\alpha }f)$, Eq \ref{eq11} itself must be invariant
under this transformation. Because the linear terms scale by $\lambda
^{\alpha }$ whereas the nonlinear diffusion terms scale by $\lambda
^{2\alpha -2\beta }$, invariance of Eq \ref{eq11} immediately fixes $\alpha
=2\beta $. Substituting Eq \ref{eq23} with $\alpha =2\beta $ into Eq \ref%
{eq11} (see Appendix \ref{A-MDD}) eliminates $\eta $ completely and reduces
Eq \ref{eq11} to an ordinary differential equation for $\Gamma (\zeta )$: by
setting $B=(1-\beta )b$, we obtain%
\begin{equation}
2b[\Gamma \frac{d^{2}\Gamma }{d\zeta ^{2}}+(1-c)(\frac{d\Gamma }{d\zeta }%
)^{2}]+\zeta \frac{d\Gamma }{d\zeta }+p\Gamma =0  \label{eq10}
\end{equation}%
with $p=\frac{A+2\beta }{1-\beta }$. In previous studies \cite%
{barenblatt2003scaling}, an equation corresponding to Eq \ref{eq10} was
derived in a quite different manner and the problem was shown to reduce to a
nonlinear eigenvalue problem as detailed in Appendix \ref{A-MDD}. Thus, we
expect a solution exists only for selected values of $p$.

In fact, for Eq \ref{eq10}, we successfully find a solution in the form $%
\Gamma (\zeta )=C(X_{f}^{2}-\zeta ^{2})$, if $p$ is given by 
\begin{equation}
p=\frac{1}{1-c},  \label{eq8}
\end{equation}%
where $X_{f}$ is a dimensionless constant representing the rescaled front
position. While $X_{f}$ remains an integration constant determined by the
specific initial profile for $c\neq 0$ owing to the lack of mass
conservation, it is uniquely fixed as $X_{f}=9^{1/3}$ in the conservative
limit $c=0$.

In this manner, we obtain, with $\zeta =\xi /\eta ^{\beta }$, 
\begin{eqnarray}
H^{\ast }(T,X,X_{0}) &=&CT^{-A/B}\eta ^{2\beta }(X_{f}^{2}-\zeta ^{2})
\label{eq3} \\
&=&C\frac{X_{f}^{2}}{T^{A/B}}\left( \frac{X_{0}}{T^{1/B}}\right) ^{2\beta
}\left( 1-\frac{X^{2}}{\widetilde{X}_{f}^{2}}\right)  \label{eq4}
\end{eqnarray}%
for $\left\vert X\right\vert <\widetilde{X}_{f}\equiv
X_{f}T^{1/B}(X_{0}/T^{1/B})^{\beta }$ and $H^{\ast }(T,X,X_{0})=0$ otherwise.

The stationary solution still leaves the exponent $B$ undetermined. To fix $%
B $, we restore dimensional variables via $h^{\ast }(t,x,l)=h_{0}H^{\ast
}(T,X,X_{0})$ with $h_{0}=m_{0}/l_{0}$, $X=x/l_{0}$, $X_{0}=l/l_{0}$, and $%
T=\kappa m_{0}t/l_{0}^{3}$. Using the scale-invariance relation $A=B-2$, the
amplitude of $h^{\ast }(t,x,l)$ and the dimensional front position $%
x_{f}=l_{0}\widetilde{X}_{f}$ carry prefactors proportional to $%
l_{0}^{2(1-\beta )(1-3/B)}$ and $l_{0}^{(1-\beta )(1-3/B)}$, respectively.
Because $l_{0}$ was introduced solely as an arbitrary unit of length,
physical observables must be independent of $l_{0}$, which requires $%
(1-\beta )(1-3/B)=0$. Combining Eq.~\eqref{eq8} with $A=B-2$ gives $%
B=(1-\beta )(3-2c)/(1-c)$, so $\beta =1$ is excluded by $B>0$. Consequently,
invariance under the choice of the arbitrary reference length $l_{0}$
uniquely selects $B=3$ (and hence $A=1$)---the same unit-choice invariance
that underpins the scale-selection mechanism in Sec~\ref{sec:fatigue}. More
generally, rather than normalizing lengths by an intrinsic characteristic
scale of the problem (such as $l$ here, or $l_{0}$ and $l_{c}$ in Sec.~\ref%
{sec:fatigue}), introducing an external arbitrary unit of length keeps all
competing physical scales on an equal footing. This avoids singling out any
scale \textit{a priori} and turns invariance under the arbitrary choice of
length unit into an exact symmetry constraint that fixes scaling exponents
and selects surviving scales.

Substituting $B=3$ and $A=1$ into Eq.~\eqref{eq8} yields the anomalous
exponent (denoted by $q$ in Ref \cite{barenblatt2003scaling}) 
\begin{equation}
\beta =\frac{c}{3-2c},
\end{equation}%
together with $C=\frac{1}{4b(1-c)}=\frac{1}{4(3-2c)}$. Thus, Eq.~\eqref{eq4}
takes the form $H^{\ast }(T,X,X_{0})$ $=$ $\frac{X_{f}^{2}}{4(3-2c)T^{1/3}}$ 
$\left( \frac{X_{0}}{T^{1/3}}\right) ^{2\beta }$ $\left( 1-\frac{X^{2}}{%
\widetilde{X}_{f}^{2}}\right) $ for $\left\vert X\right\vert <\widetilde{X}%
_{f}$, i.e., a memory-retaining II fixed point with $\alpha =2\beta $, which
continuously recovers the conservative DDD fixed point, i.e., the
memory-loss fixed point, in Eq.~\eqref{eq6} in the limit $c\rightarrow 0$ ($%
\beta \rightarrow 0$). In dimensional form, for $\left\vert x\right\vert
<x_{f}\equiv X_{f}(\kappa m_{0}t)^{\frac{1-\beta }{3}}l^{\beta }$, it reads 
\begin{equation}
h^{\ast }(t,x,l)=\frac{X_{f}^{2}m_{0}^{\frac{2(1-\beta )}{3}}l^{2\beta }}{%
4(3-2c)(\kappa t)^{\frac{1+2\beta }{3}}}\left[ 1-\left( \frac{x}{%
X_{f}(\kappa m_{0}t)^{\frac{1-\beta }{3}}l^{\beta }}\right) ^{2}\right] ,
\label{eq22}
\end{equation}%
which reproduces a known result in a different manner, thereby providing
another nontrivial validation of the unified RG formulation.

In contrast with Eq \ref{eq6a}, Eq \ref{eq22} is dependent on both $m_{0}$
and $l$: it depends on $m_{0}$ and $l$ solely through the combination $%
m_{0}^{(1-\beta )/3}l^{\beta }$, or equivalently $m_{0}l^{\frac{3\beta }{%
1-\beta }}$ with the anomalous dimension $\beta $. This factor originates
from the expression $f(\xi ,\eta )=\eta ^{2\beta }\Gamma (\xi /\eta ^{\beta
})$ with $\xi =x/(\kappa m_{0}t)^{1/3}$ and $\eta =l/(\kappa m_{0}t)^{1/3}$,
for which long-time behavior corresponds to $\eta \rightarrow 0$. This limit
is achieved in two separate ways: (I) $t\rightarrow \infty $ for a fixed $l$%
, and (II) $l\rightarrow 0$ for a fixed $t$. However, the factor $m_{0}l^{%
\frac{3\beta }{1-\beta }}$ goes to zero or infinity depending on the sign of
the exponent in the limit II, if $m_{0}$ is independent of $l$, which
contradicts with the result of limit I. This argument concludes that $m_{0}$
does depend on $l$ such that $m_{0}l^{\frac{3\beta }{1-\beta }}=const$.

For small $\beta $, the factor $m_{0}l^{\frac{3\beta }{1-\beta }}$ appearing
in Eq \ref{eq22} can be expanded as $m_{0}l^{\frac{3\beta }{1-\beta }}$ $=$ $%
m_{0}$ $(1+3\beta \log l+\cdots )$ . The resulting logarithmic contribution
originates from the surviving relevant scale $l$ and therefore provides a
direct measure of information retained under RG flow. This observation also
clarifies the connection with field-theoretic RG: the logarithmic dependence
is reminiscent of the logarithmic divergences appearing in the Barenblatt
problem. The divergent factor may be absorbed by introducing a
renormalization constant as in QFT \cite{ryder1996quantum}, where $m_{0}$
and $l$ play roles analogous to the \textit{bare mass} and \textit{%
ultraviolet cutoff}, respectively. In this sense, anomalous dimensions may
be interpreted as signatures of information retention associated with
relevant scales that survive RG flow.

\section{Three-Scale Extension: Barenblatt's Fatigue-Crack Growth Problem}

\label{sec:fatigue}

An advantage of the present RG approach over conventional similarity methods
and the dynamical-system description (DSD) is that it does not assume
self-similarity a priori; instead, self-similar solutions emerge naturally
from scale invariance of governing equations \cite{Okumura2026nonlinear}. To
demonstrate that this principle extends to self-similarity of the second
kind even when the governing equation is not completely specified---and it
simultaneously determines which relevant scales survive when more than two
length scales compete---we examine Barenblatt's incomplete-similarity
problem of fatigue crack growth \cite{Barenblatt, barenblatt2014flow}.

Under cyclic tensile loading with stress amplitude $\Delta \sigma =\sigma
_{\max }-\sigma _{\min }$ and cycle asymmetry ratio $R=\sigma _{\max
}/\sigma _{\min }$, dimensional analysis of the cycle-averaged growth rate $%
dl/dn$ of a crack of length $l(n)$ in a specimen of characteristic size $h$
yields \cite{barenblatt2014flow} 
\begin{equation}
\frac{dl}{dn}=\varepsilon ^{2}l\,\Phi \!\left( \varepsilon \sqrt{\frac{l}{%
l_{c}}},R,\sqrt{\frac{h}{l_{c}}}\right) ,  \label{fatigue_ode}
\end{equation}%
where $\varepsilon \equiv c\Delta \sigma /\sigma _{Y}$ is the dimensionless
stress amplitude (with geometric factor $c$ and yield stress $\sigma _{Y}$), 
$l_{c}\equiv (K_{Ic}/\sigma _{Y})^{2}$ is the Irwin microscopic
fracture-process-zone scale (with fracture toughness $K_{Ic}$), and $\Phi $
is an unspecified dimensionless function. Together with the initial
condition $l_{0}$, three independent physical length scales compete in the
problem: the initial crack length $l_{0}$, the microscopic fracture scale $%
l_{c}$, and the macroscopic specimen size $h$.

When the crack growth dynamics exhibits finite-time blow up at a critical
cycle number $n_{f}$, introducing an external arbitrary reference length $l_{%
\mathrm{ref}}$ (rather than normalizing by an intrinsic scale such as $l_{0}$
or $l_{c}$, as in Sec.~IV), the dimensionless crack-length field $%
H(T;X_{0},X_{1},X_{2})=l(n)/l_{\mathrm{ref}}$, the three dimensionless
relevant scales $X_{0}=l_{0}/l_{\mathrm{ref}}$, $X_{1}=l_{c}/l_{\mathrm{ref}%
} $, $X_{2}=h/l_{\mathrm{ref}}$, and the dimensionless remaining cycles to
failure $T=(n_{f}-n)/t_{\mathrm{ref}}>0$ with $1/t_{\mathrm{ref}}=\epsilon
^{2}$, Eq \ref{fatigue_ode} takes the form of Eq \ref{eq13}:

\begin{equation}
\partial _{T}H(T;X_{0},X_{1},X_{2})=-H\,\Phi \!\left( \varepsilon \sqrt{%
\frac{H}{X_{1}}},R,\sqrt{\frac{X_{2}}{X_{1}}}\right) .
\label{fatigue_rg_pde}
\end{equation}%
Here, treating the discrete cycle number $n$ as a continuous variable is
justified in the multi-cycle fatigue regime $\varepsilon \ll 1$, where the
per-cycle increment $\Delta T=1/t_{\mathrm{ref}}=\varepsilon ^{2}\ll 1$ is
small; accordingly, the $T\rightarrow 0^{+}$ limit describes the
intermediate asymptotic regime $\varepsilon ^{2}\ll T\ll 1$ ($1\ll
n_{f}-n\ll n_{f}$). Because this dynamics approaches blow-up as $%
T\rightarrow 0^{+}$ ($n\rightarrow n_{f}$), it belongs to Case~I ($0<L<1$)
of the BKL-type scale transformations in Eqs.~\eqref{eq14}--\eqref{eq16}.
Note that although Eq \ref{fatigue_rg_pde} contains no spatial derivatives
(as in the RG formulation of bubble breakup \cite{Okumura2025RG}), it is a
partial differential equation of the form of Eq \ref{eq13} for the field $%
H(T;X_{0},X_{1},X_{2})$ defined on the extended state space $%
(X_{0},X_{1},X_{2})$.

For $H$ governed by Eq \ref{fatigue_rg_pde}, the RG framework predicts
self-similar RG fixed points of the form $H^{\ast
}(T;X_{0},X_{1},X_{2})=T^{-A/B}f(\eta _{0},\eta _{1},\eta _{2})$, where $%
\eta _{i}=X_{i}/T^{1/B}$ ($i=0,1,2$) if the solution $f$ to the RG
fixed-point equation exists and if Eq \ref{fatigue_rg_pde} flows towards a
fixed point, where Eq \ref{fatigue_rg_pde} becomes scale invariant. Applying
rescaling $(T,\mathbf{X})\rightarrow (L^{B}T,L\mathbf{X})$ with $H(L^{B}T,L%
\mathbf{X})=H_{L}(T,\mathbf{X})/L^{A}$ and requiring the scale invariance of
Eq \ref{fatigue_rg_pde}, we conclude that $\Phi (u,R,Z)$ should be a
power-law function of $u$, and the exponent $\alpha (R,Z)$ should satisfy $B=%
\frac{1}{2}\alpha (R,Z)(A+1)$. This is because, while $R$ and $Z$ are
unchanged by the rescaling, $u$ is changed to $L^{-(A+1)/2}u_{L}$. Although
the right-hand side of Eq \ref{fatigue_rg_pde} can contain irrelevant terms
as discussed in Appendix~\ref{app:fatigue}, in the following we neglect them
to investigate asymptotic states, and thus we obtain $\Phi (u,R,Z)\simeq
u^{\alpha (R,Z)}\Phi _{1}(R,Z)$. In other words, the $H$ dependence of the
right-hand side of the governing equation in Eq \ref{fatigue_rg_pde} is
specified as $H^{1+\alpha /2}$, i.e., $F(H,\cdots )$ in Eq \ref{eq13} is
identified with $-H(\varepsilon \sqrt{H/X_{1}})^{\alpha (R,Z)}\Phi _{1}(R,Z)$%
. In this case, the RG fixed-point equation for $f(\eta _{0},\eta _{1},\eta
_{2})$ is given by 
\begin{equation}
0=Af-Bf(\varepsilon \sqrt{f/\eta _{1}})^{\alpha (R,Z)}\Phi
_{1}(R,Z)+\sum_{i=0}^{2}\eta _{i}\frac{\partial f}{\partial \eta _{i}},
\label{fatigue_fp_eq}
\end{equation}%
with $Z=\sqrt{\eta _{2}/\eta _{1}}$.

As shown in Appendix~\ref{app:fatigue}, applying the same $\lambda $%
-rescaling invariance method as in Sec \ref{sec:mddd}---here induced by
unit-choice invariance under changes of the arbitrary reference length $l_{%
\mathrm{ref}}$---we find 
\begin{equation}
f(\eta _{0},\eta _{1},\eta _{2})=\eta _{1}^{\widetilde{\alpha }}\Gamma (\eta
_{2}/\eta _{1}^{\beta })\text{ \ (}\widetilde{\alpha }=1,\ \beta =1),
\label{eqA}
\end{equation}%
in direct analogy with Section~\ref{sec:mddd}. Here, we have set $\eta
_{1}=X_{1}/T^{1/B}$ and $\eta _{2}=X_{2}/T^{1/B}$. In the asymptotic
solution in Eq \ref{eqA}, the stationary RG flow equation eliminates the
initial scale $\eta _{0}(=l_{0}/l_{\mathrm{ref}})$ while retaining the two
competing scales $\eta _{1}$ and $\eta _{2}$. More specifically, $\eta _{1}$
appears as the multiplicative power-law prefactor ($\widetilde{\alpha }=1$)
and rescales the profile argument ($\beta =1$), while the macroscopic
specimen scale $\eta _{2}=X_{2}/T^{1/B}$ enters solely in the numerator of
the profile ratio. More explicitly, we find $\Gamma \left( \eta _{2}/\eta
_{1}=Z^{2}\right) =\left( \frac{2}{\alpha (R,Z)\,\Phi _{1}(R,Z)\,\varepsilon
^{\alpha (R,Z)}\,}\right) ^{2/\alpha (R,Z)}$. By noting $H^{\ast }=T^{-A/B}f$
and $T^{-A/B}\eta _{1}=T^{-(A+1)/B}X_{1}=T^{-2/\alpha (R,Z)}X_{1}$---where
the individual arbitrariness in $A$ and $B$ cancels out via the
scale-invariance relation $B=\frac{1}{2}\alpha (R,Z)(A+1)$---we obtain 
\begin{equation}
H^{\ast }(T;X_{0},X_{1},X_{2})=X_{1}\left( \frac{2}{\alpha (R,Z)\Phi
_{1}(R,Z)\epsilon ^{\alpha (R,Z)}T}\right) ^{2/\alpha (R,Z)}.
\label{fatigue_Hstar}
\end{equation}%
As shown in Appendix~\ref{app:fatigue}, this solution recovers Barenblatt's
Paris scaling law \cite{barenblatt2014flow}.

While Paris's law follows algebraically once scale invariance selects $\Phi
(u,R,Z)\simeq u^{\alpha (R,Z)}\Phi _{1}(R,Z)$, formulating the asymptotic
dynamics as an RG fixed point in the three-scale space $(X_{0},X_{1},X_{2})$
serves two essential conceptual purposes. First, it identifies the
asymptotic solution underlying Paris's law directly as a scale-invariant RG
fixed point toward which equations with irrelevant deviations flow. Second,
and more fundamentally, it demonstrates how the RG fixed-point concept
itself is extended in multi-scale problems: starting from $%
(X_{0},X_{1},X_{2})$, the stationary RG flow equation eliminates $X_{0}$,
retains $X_{1}$ as an explicit amplitude scale ($\eta _{1}$), and preserves $%
X_{2}$ ($\eta _{2}$) through the locked similarity variable $\eta _{2}/\eta
_{1}^{\beta }=Z^{2}$ inside $\Phi _{1}(R,Z)$ and $\alpha (R,Z)$. This
realizes the spatial-free counterpart of a memory-retaining II fixed point
(with nonzero $\alpha $ and $\beta $), from which progressive scale
decoupling can be traced explicitly:

\begin{itemize}
\item For finite specimen size $Z=\sqrt{h/l_{c}}$ with $\alpha (R,Z)\neq 0$,
both $\eta _{1}$ ($X_{1}$) and $\eta _{2}$ ($X_{2}$) survive the RG flow and
remain encoded in both the anomalous amplitude and the $Z$-dependent scaling
function $\Phi _{1}(R,Z)$ and exponent $\alpha (R,Z)$.

\item In the infinite-specimen limit $Z\rightarrow \infty $ ($h\gg l_{c}$)
with $\alpha (R,\infty )\neq 0$ and finite $\Phi _{1}(R,\infty )$, $\Gamma
(Z^{2})$ approaches a constant $\Gamma (\infty )$ so that the macroscopic
scale $\eta _{2}$ ($X_{2}$) decouples, leaving only the single microscopic
scale $\eta _{1}$ ($X_{1}$) (i.e., $l_{c}$).

\item When $\alpha (R,Z)\rightarrow 0$ as $Z\rightarrow \infty $, all three
scales $(X_{0},X_{1},X_{2})$ drop out (corresponding to the classical Paris
exponent $2+\alpha =2$), recovering the \textit{memory-loss} regime.
\end{itemize}

Thus, memory retention is not restricted to initial-condition length scales:
material scales ($l_{c}$), geometric scales ($h$), and loading parameters ($%
R $) are governed by the exact same RG fixed-point selection mechanism. Even
when the underlying microscopic constitutive function $\Phi $ contained in
the governing equation is unspecified---so that $\alpha (R,Z)$ cannot be
evaluated analytically---the stationary RG flow equation still determines,
without assuming self-similarity a priori, both the emergence of an
incomplete-similarity fixed point and its canonical memory-retaining
structure $\eta _{1}^{\tilde{\alpha}}\Gamma (\eta _{2}/\eta _{1}^{\beta })$.

\section{\noindent Discussion}

\label{sec:discussion}

\subsection{Symmetry principles and universality classes}

The present work extends the predictive RG framework of Ref \cite%
{Okumura2026nonlinear} by introducing a new mode of application in which all
competing physical length scales---such as initial, material, and geometric
scales---are retained on an equal footing inside the extended state vector $%
\mathbf{X}$, rather than restricting $\mathbf{X}$ to the spatial coordinate
alone from the outset. While retaining these auxiliary scales opens
additional scaling degrees of freedom in the stationary RG flow equation,
combining two complementary symmetry principles closes the problem
non-perturbatively without an \textit{a priori} similarity ansatz: (i) the $%
\lambda $-rescaling invariance method applied to the stationary RG flow
equation, which reduces the multi-variable flow equation to an ordinary
differential equation without assuming separation of variables and
constrains the anomalous exponents $(\alpha ,\beta )$, and (ii) unit-choice
invariance under changes of the external arbitrary reference length ($l_{0}$
in Secs \ref{sec:ddd}--\ref{sec:mddd} and $l_{\mathrm{ref}}$ in Sec \ref%
{sec:fatigue}), which keeps all competing scales unbiased and fixes the
remaining scaling dimensions $(A,B)$ (or their invariant ratio) while
selecting which physical scales survive. In doing so, the present
formulation provides a systematic prescription for treating general
second-kind self-similar solutions with $\beta \neq 0$ within the RG
framework---a regime that cannot be formulated when $\mathbf{X}$ is
restricted to a single spatial coordinate.

The modified DDD model therefore illustrates a representative example in
which anomalous scaling emerges together with explicit retention of a
relevant length scale. The framework predicts that the\textit{\ universality
class} for the modified DDD model consists of PDEs of the form $\partial
_{T}H\mathbf{(}T,X)=$ $2[\partial _{X}(H\partial _{X}H)-c(\partial
_{X}H)^{2}]$ $+G$, where $G$ is given by a linear combination of terms that
can be expressed as $H^{n_{0}}(\partial _{X}H)^{n_{1}}(\partial
_{X}^{2}H)^{n_{2}}(\partial _{X}^{3}H)^{n_{3}}\cdots $ with a \textit{%
negative} $M$ in Eq \ref{M1} with $B=3$ and $A=1$. Remarkably, the
irrelevant condition $M<0$ is independent of the anomalous dimension $\beta $%
. The anomalous dimension therefore governs the asymptotic memory of the
initial scale without affecting the RG classification of relevant and
irrelevant terms.

Barenblatt's fatigue-crack growth problem (Sec \ref{sec:fatigue}) extends
this interplay between universality classes and memory retention to a
multi-scale setting in which the governing equation is not completely
specified a priori. Here, the RG framework predicts a universality class of
blow-up dynamics (Case I, $0<L<1$) of the form $\partial _{T}H=$ $%
-\varepsilon ^{\alpha (R,Z)}$ $\Phi _{1}(R,Z)$ $X_{1}^{-\alpha (R,Z)/2}$ $%
H^{1+\alpha (R,Z)/2}$ $+G$, where $G$ consists of lower-order subleading
terms proportional to $X_{1}^{-k/2}$ $H^{1+k/2}$. These subleading terms
carry scaling exponents $M_{k}=\frac{1}{2}(A+1)[\alpha (R,Z)-k]$ where $%
A+1>0 $ (see Appendix \ref{app:fatigue}). In contrast to the modified DDD
model---where the irrelevance criterion $M<0$ is completely independent of
the anomalous exponent $\beta $---here the irrelevance condition $M_{k}>0$
reduces to $k<\alpha (R,Z)$ and thus depends directly on the anomalous
exponent $\alpha (R,Z)$ (and hence on $R$ and $Z=\sqrt{h/l_{c}}$). Moreover,
because $l_{\mathrm{ref}}$ does not enter $T$, $A$ and $B$ remain
individually arbitrary while their invariant ratio is fixed as $%
(A+1)/B=2/\alpha (R,Z)$ (see Appendix \ref{app:fatigue}); consequently, the
anomalous exponent $\alpha (R,Z)$ governs both the RG classification of
irrelevant terms and the invariant ratio of scaling dimensions.

\subsection{Unified classification of memory-loss and memory-retaining fixed
points}

From the RG point of view, the central result of the present work is not the
classification itself but the extension of fixed-point theory from
single-variable profiles to fixed-point functions that can retain explicit
dependence on asymptotically relevant scales. Conventional PDE RG
formulations accommodate anomalous amplitudes ($\beta =0$), but the present
extended-state-space formulation also admits profile-reshaping fixed points (%
$\beta \neq 0$), thereby providing a general RG framework for
self-similarity of the second kind.

The unified framework also encompasses the $\beta =0$ class of second-kind
self-similarity (memory-retaining I), as illustrated by Barenblatt's
equation, where the diffusion coefficient takes two distinct constant values
depending on the sign of $\partial _{t}h$ and thereby breaks mass
conservation \cite{Barenblatt,Goldenfeld}. Notably, in the previous study~%
\cite{okumura2026combined, Okumura2026nonlinear}, with the single-variable
identification $\mathbf{H}(T;\mathbf{X})=H(T,X)$, the RG framework recovered
the known scaling solution without retaining $X_{0}$ explicitly in the state
vector. From a unified perspective, however, we should retain all relevant
length scales by setting $\mathbf{H}(T;\mathbf{X})=H(T;X,X_{0})$. In this
extended representation, applying the same $\lambda $-rescaling invariance
method to the stationary RG flow equation directly yields the Barenblatt
solution: matching the $\lambda $-rescaling of the terms in the fixed-point
equation forces $\beta =0$ and fixes the amplitude exponent to $2\alpha $
(with $\alpha $ denoting Barenblatt's anomalous exponent; see Appendix \ref%
{A-BD}), yielding $f(\xi ,\eta )=\eta ^{2\alpha }\Gamma (\xi )$. The
resulting $\beta =0$ (memory-retaining I) solution, proportional to a
divergent series $m_{0}$ $(1+2\alpha \log l+\cdots )\simeq m_{0}l^{2\alpha }$
for small $l$, reproduces the known result through a completely different
route, thereby providing an additional nontrivial validation of the unified
RG formulation. The logarithmic dependence appearing in the Barenblatt case
is closely analogous to that obtained above for the modified DDD model. In
both examples, the logarithmic term originates from a relevant scale that
survives RG flow and is subsequently encoded in the asymptotic fixed-point
function through an anomalous dimension. Crucially, however, whereas the $%
\beta =0$ class could still be accommodated within the single-variable
identification $\mathbf{H}(T;\mathbf{X})=H(T,X)$ because the auxiliary scale
enters solely as an overall amplitude prefactor, the general $\beta \neq 0$
class involves an explicit coupling between the spatial coordinate and
auxiliary physical scales inside the similarity variable $\xi /\eta ^{\beta
} $ (or $\eta _{2}/\eta _{1}^{\beta }$). Retaining physical length scales
inside the state vector $\mathbf{X}$ is therefore indispensable for
constructing a closed RG theory whenever $\beta \neq 0$.

The unified framework can also recover a known result in principle for
mass-conserving diffusion models such as linear diffusion and DDD with the
identification $\mathbf{H}(T;\mathbf{X})=H(T;X,X_{0})$ (see Appendix \ref%
{A-DD}), where the asymptotic self-similar solution depends on the conserved
quantity $m_{0}${} but not on the initial length scale $l$ (see e.g., Eq \ref%
{eq6a}). By contrast, in the nonconservative Barenblatt and modified DDD
models, the initial scale $l$ survives RG evolution and remains encoded in
the asymptotic state (see Eq \ref{eq22}), while in the multi-scale
fatigue-crack problem the initial crack scale $\eta _{0}$ is forgotten
whereas the material and geometric scales $\eta _{1}$ and $\eta _{2}$
survive (see Eq \ref{eqA}). The conservative and multi-scale FPs therefore
exhibit a stronger form of universality, while the nonconservative FPs
retain additional information about the surviving physical scales.

The examples of conservative diffusion (linear diffusion and DDD),
Barenblatt diffusion, the modified DDD model, and fatigue-crack growth
suggest that the vector $\mathbf{X}$ appearing in the unified RG framework
may be usefully interpreted as containing all relevant length scales in the
problem, thereby allowing memory loss and memory retention to be treated on
an equal footing. From this viewpoint, the asymptotic fixed point is not
assumed a priori to depend only on conventional similarity variables.
Instead, whether a given relevant scale survives or is eliminated under
repeated RG transformations emerges from the RG flow itself. This viewpoint
naturally motivates the meaning of memory-retaining fixed points.

Unlike Hamiltonian-based RG, where surviving scales parameterize fixed
manifolds in coupling space, the present PDE formulation encodes surviving
scales directly as explicit arguments of the asymptotic fixed-point function 
$f$, thereby allowing fixed points of the form $\eta ^{\alpha }\Gamma (\xi
/\eta ^{\beta })$---a distinction that concerns the functional structure of
the fixed point rather than anomalous scaling alone. The memory-based
classification proposed here should therefore be viewed not as an
independent assumption, but as a natural consequence of allowing RG fixed
points to retain asymptotically relevant physical scales. Taken together,
the examples analyzed above naturally organize self-similar RG fixed points
into three classes: (i) \textit{memory-loss} ($\alpha =\beta =0$), where
auxiliary scales drop out of $f(\xi ,\eta )=\Gamma (\xi )$, realized in
conservative diffusion (linear diffusion and DDD); (ii) \textit{%
memory-retaining I} ($\alpha \neq 0$, $\beta =0$), where $f(\xi ,\eta )=\eta
^{\alpha }\Gamma (\xi )$ retains auxiliary scales solely in the
multiplicative amplitude, as in Barenblatt diffusion; and (iii) \textit{%
memory-retaining II} ($\alpha \neq 0$, $\beta \neq 0$), where surviving
scales also reshape the similarity argument, realized in the modified DDD
model via $f(\xi ,\eta )=\eta ^{\alpha }\Gamma (\xi /\eta ^{\beta })$.

Remarkably, the three-scale fatigue-crack problem spans the spatial-free
counterparts of all three classes across its asymptotic regimes (with $\eta
_{0}$ always eliminated): the classical Paris limit ($\alpha \rightarrow 0$)
recovers the memory-loss class; the infinite-specimen limit ($Z\rightarrow
\infty $ with $\alpha (R,\infty )\neq 0$) reduces to memory-retaining I as $%
\eta _{2}$ decouples and only $\eta _{1}^{\widetilde{\alpha }}$ survives;
and the finite-specimen regime ($Z<\infty $ with $\alpha (R,Z)\neq 0$)
realizes memory-retaining II via $\eta _{1}^{\widetilde{\alpha }}\Gamma
(\eta _{2}/\eta _{1}^{\beta })$.

Interestingly, memory loss, partial memory, and memory retention have all
been observed experimentally in self-similar hydrodynamic singularities.
Both memory loss and memory retention were identified in fluid-drop
pinch-off dynamics \cite{1994ScienceNagelDropFallingFaucet,
2003ScienceNagelMemoryDropBreakup, pahlavan2019restoring}, where the
asymptotic behavior becomes independent of or dependent on the initial
conditions, respectively. Partial memory was introduced more recently in
confined bubble breakup \cite{nakazato2018self, yoshino2025partial}, where a
finite amount of information survives asymptotically. However, a unified RG
understanding of these distinct memory regimes has remained unclear.

The present work offers a common RG perspective on these phenomena. Together
with previous RG studies of memory loss in universal singular dynamics \cite%
{1993PRLEggersPinchoff, 1994ScienceNagelDropFallingFaucet} by the unified RG
framework \cite{Okumura2026nonlinear}, our examples show that RG evolution
can lead to complete elimination, partial retention, or persistent survival
of initial-condition information. Thus, the central question is not merely
whether dynamics are universal, but which information survives repeated RG
transformations and how it is encoded in the asymptotic FP.

Recent applications of the same unified RG framework have revealed
additional forms of RG behavior beyond conventional universality classes.
These include traveling-wave fixed points that do not generate universality
classes \cite{Okumura2026traveling} and front-selection mechanisms that are
not determined by RG fixed-point stability \cite{Okumura2026FKPP}.

Together with the memory-retaining fixed points introduced here, these
examples show that RG fixed points in nonlinear PDEs possess a richer
structure than the standard paradigm of universality alone. From this
viewpoint, the role of an RG fixed point extends beyond organizing
universality classes: universality, memory retention, and asymptotic-state
selection represent distinct aspects of RG behavior, corresponding
respectively to information loss, information retention, and state selection
under RG flow.

\section{\noindent Conclusion}

\label{sec:conclusion}

In this work, we have established a unified, non-perturbative
renormalization-group (RG) formulation capable of treating general
second-kind self-similar dynamics---including the $\beta \neq 0$ regime that
could not be formulated within conventional single-variable PDE RG
approaches. By incorporating all competing physical length scales (such as
initial, material, and geometric scales) on an equal footing into the
extended state vector $\mathbf{X}$, and by combining unit-choice invariance
under an arbitrary external reference length with the $\lambda $-rescaling
invariance method applied to the stationary RG flow equation, our framework
treats first-kind self-similarity ($\alpha =\beta =0$) as well as both $%
\beta =0$ and $\beta \neq 0$ classes of second-kind self-similarity within a
single, systematic prescription.

From the perspective of self-similar analysis, a key contribution of this
study is providing concrete demonstrations in which second-kind self-similar
solutions---and the asymptotic selection of relevant length scales---emerge
self-consistently from the RG flow without assuming an a priori similarity
ansatz. Across conservative linear and density-dependent diffusion, the RG
flow eliminates the initial scale to recover the classical first-kind
scaling; in nonconservative diffusion, it preserves the initial scale either
purely in the amplitude for Barenblatt's equation ($\beta =0$) or by
reshaping both the amplitude and the similarity variable for the modified
density-dependent diffusion model ($\beta \neq 0$); and in Barenblatt's
three-scale fatigue-crack growth problem, it eliminates the initial crack
length while combining the microscopic fracture scale and specimen size into
a spatial-free $\beta \neq 0$ fixed point that recovers Paris's scaling law
even though the governing equation is not completely specified.

Conceptually, retaining physical scales inside $\mathbf{X}$ broadens the
notion of an RG fixed point in nonlinear PDEs: rather than depending solely
on conventional similarity variables, the asymptotic fixed-point function $f$
can explicitly retain surviving physical scales as arguments. This unified
structure naturally leads to a memory-based classification of self-similar
RG fixed points---comprising memory-loss ($\alpha =\beta =0$),
amplitude-only memory-retaining I ($\alpha \neq 0$, $\beta =0$), and
profile-reshaping memory-retaining II ($\alpha \neq 0$, $\beta \neq 0$)
fixed points---expressed canonically as $f(\xi ,\boldsymbol{\eta })$ $=\eta
^{\alpha }\Gamma (\xi /\eta ^{\beta })$ (or its spatial-free counterpart $%
\eta _{1}^{\alpha }\Gamma (\eta _{2}/\eta _{1}^{\beta })$).

More broadly, whereas Hamiltonian-based RG characterizes critical behavior
through fixed points and fixed manifolds in coupling space, RG formulations
acting directly on dynamical fields in nonlinear PDEs reveal how
physical-scale information is either erased or encoded in the asymptotic
fixed-point function itself. By shifting the focus from determining scaling
exponents alone to classifying the information that survives repeated RG
transformations, memory-retaining fixed points provide a general organizing
principle for self-similar and multi-scale nonlinear dynamics.

\textit{The author is grateful to Professor Nigel Goldenfeld (UCSD) for
helpful comments and encouragement. This work was supported by JSPS\ KAKENHI
Grant Number JP24K00596. }\newline

\section*{Declarations}

\section*{Conflict of Interest}

The author declares that there is no conflict of interest.

\section*{Data Availability}

Data sharing is not applicable to this article as no new data were created
or analyzed in this study.

\appendix
\clearpage

\section{Supplementary notes\label{A2}}

\subsection{Note on the unified RG framework\label{A-U}}

For completeness, we summarize the minimum ingredients of the unified RG
framework needed in the present work, so that the discussion below is
self-contained. The term in $F$ and $G$ in Eq \ref{eq13} can be generally
written as%
\begin{equation}
H^{N_{0}}\cdots (\partial _{X}H)^{N_{1}}(\partial _{X}^{2}H)^{N_{2}}\cdots ,
\label{t1}
\end{equation}%
which, under the scale transformation, is changed into the form $%
L^{M}(H_{L})^{N_{0}}(\partial _{X}H_{L})^{N_{1}}(\partial
_{X}^{2}H_{L})^{N_{2}}\cdots $, with the \textit{scaling factor} $L^{M}$
characterized by the \textit{scaling exponent} $M$, which defines relevance
of the term:%
\begin{equation}
M=A+B-[N_{0}A+N_{1}(A+1)+N_{2}(A+2)+\cdots ],  \label{M1}
\end{equation}

We are interested in \textit{singular dynamics} (Case I) or \textit{%
intermediate asymptotics} (Case II) with the boundary condition at $T=1$: $%
H(1,\mathbf{X})=h(\mathbf{X})$. The RG transformation is defined for $h(%
\mathbf{X})$ as%
\begin{equation}
\emph{R}_{L,\mathbf{G}}h(\mathbf{X})\equiv L^{A}H(L^{B};L\mathbf{X})=H_{L}(1;%
\mathbf{X})\text{.}  \label{eq17}
\end{equation}%
The second equality is based on Eq \ref{eq16}.

Repeated application of the RG transformation generates a flow in function
space. A fixed point corresponds to a profile invariant under further RG
transformations. Universality classes arise when different initial
conditions flow toward the same fixed point. If $F$ is scale-invariant and $%
G $ is irrelevant, as assumed, we can expect that iteration of RG makes $h(%
\mathbf{X})$ and Eq \ref{eq13} flow into their FPs: $f(\mathbf{X})$ and 'Eq %
\ref{eq13} with zero $G$,' where the FP is defined by the following
equation: 
\begin{equation}
\emph{R}_{L,\mathbf{G}^{\ast }}f(\mathbf{X})=f(\mathbf{X})\Leftrightarrow
L^{A}H^{\ast }(L^{B};L\mathbf{X})=f(\mathbf{X}).
\end{equation}%
If such a point exists, setting $T=L^{B}$ in this equation results in $%
T^{A/B}H(T;T^{1/B}\mathbf{X})=f(\mathbf{X})$, from which we obtain Eq \ref%
{e20}: $H^{\ast }(T;\mathbf{X})=T^{-A/B}f(\mathbf{X}/T^{1/B})$.

The RG flow equation can be obtained for $\widetilde{f}(\tau ;\mathbf{X}%
)\equiv H_{L}(1;\mathbf{X})=\emph{R}_{L}\mathbf{f}(\mathbf{X})$ by
introducing the logarithmic time $\tau =-B\log L$ ($\tau =B\log L$) for Case
I (Case II):

\begin{eqnarray}
\mp B\frac{d\widetilde{f}(\tau ;\mathbf{X})}{d\tau } &=&A\widetilde{f}(\tau ;%
\mathbf{X})+BF(\widetilde{f},\partial _{X}\widetilde{f},\cdots )  \notag \\
&&+\sum_{i}X_{i}\frac{\partial \widetilde{f}(\tau ;\mathbf{X})}{\partial
X_{i}},  \label{eq21}
\end{eqnarray}%
where the minus and plus signs on the left-hand side corresponds to Case I
and II, respectively. By setting $\frac{d\widetilde{f}(\tau ;\mathbf{X})}{%
d\tau }=0$ in Eq \ref{eq21}, we obtain Eq \ref{eq21A} for the stationary
solution $f(\mathbf{X})$. Its stability can be examined by substituting 
\begin{equation}
\widetilde{f}(\tau ;\mathbf{X})=f(\mathbf{X})+\bm{\delta}(\mathbf{X}%
)e^{\omega \tau }
\end{equation}%
into Eq \ref{eq21}, linearizing the equation in terms of $\bm{\delta}(%
\mathbf{X})$, and examining the sign of $\omega $. Since $\tau $ goes to
positive infinity for both Cases I and II with repeated application of RG
transformation, a negative $\omega $ corresponds to a \textit{stable}
(decaying) mode, whereas a positive $\omega $ indicates an \textit{unstable}
(growing) mode. The case $\omega =0$ corresponds to a \textit{marginal}
mode, which may be absorbed into the FP solution depending on the structure
of the problem.

\subsection{Note on the DDD model\label{A-DD}}

In the light of Eq \ref{t1}, for the two terms, $(\partial _{X}H)^{2}$ and $%
H\partial _{X}^{2}H$, respectively with $(N_{1},N_{11})=(0,2)$ and $%
(N_{1},N_{111})=(1,1)$, originating from the diffusion term $\partial
_{X}(H\partial _{X}H)$, the dimensions are respectively given as $%
M=A+B-2(A+1)$ and $A+B-A-(A+2)$, both reducing to the same value $-A+B-2$.
To ensure the scale invariance of both terms, we require $B=A+2$. If we
further require the invariance of the mass conservation law, for which $\int
dXH\rightarrow $ $\int dLXH(L^{B}T,LX)=L^{1-A}\int dXH$, we obtain $A=1$,
i.e., $B=3$.

The RG flow equation in the present case is given as

\begin{equation}
3\frac{d\widetilde{f}(\tau ,X)}{d\tau }=\widetilde{f}(\tau ,X)+6\partial
_{X}(\widetilde{f}\partial _{X}\widetilde{f})+X\frac{\partial \widetilde{f}%
(\tau ,X)}{\partial X}.
\end{equation}%
This equation is obtained in a quite different manner in previous studies 
\cite{Barenblatt}. To determine the stationary FP solution $\widetilde{f}%
(\tau ,X)=f(X)$, we set the right-hand side to zero, which can be integrated
to $\partial _{X}f^{2}+Xf/3=cnst$, where the constant should be zero because
the solution must be finite and symmetric: $f^{\prime }(0)=0$. The last
differential equation gives physical solution: $f(X)=(X_{f}^{2}-X^{2})/12$
for $\left\vert X\right\vert <X_{f}$ and $f(X)=0$ otherwise.$\ $This leads
to $H^{\ast }(T,X)$ $=T^{-1/3}(X_{f}^{2}-(X/T^{1/3})^{2})/12$, which means
the time-dependent front of the flow is given by $X=X_{f}T^{1/3}$ and thus
the mass conservation is expressed as $(1/6)%
\int_{0}^{X_{f}T^{1/3}}T^{-1/3}(X_{f}^{2}-(X/T^{1/3})^{2})dX=1$. The last
relation gives $X_{f}=9^{1/3}$. In this way, we obtain Eqs \ref{eq5} and \ref%
{eq6}.

The mass conservation is guaranteed by Eq \ref{DDD} for the solution in Eq %
\ref{eq5}, since the integration of the flux $\int_{-\infty }^{\infty
}\partial _{T}H(T,X)dX$ scales, from $\partial _{X}(H\partial _{X}H)\sim
\partial _{X}(\partial _{X}H^{2})$, as $[\partial
_{X}H^{2}]_{X=-X_{f}T^{1/3}}^{X=X_{f}T^{1/3}}$, which is zero.

If we instead start with the identification $H(T;\mathbf{X})=H(T;X,X_{0})$,
the stationary RG flow equation takes the following form: $f(\xi ,\eta
)+6\partial _{\xi }(f\partial _{\xi }f)+\xi \frac{\partial f}{\partial \xi }%
+\eta \frac{\partial f}{\partial \eta }=0$. Applying the $\lambda $%
-rescaling invariance method introduced in Sec \ref{sec:mddd} yields Eq \ref%
{eq23} and---because the linear terms and the nonlinear diffusion term scale
as $\lambda ^{\alpha }$ and $\lambda ^{2\alpha -2\beta }$,
respectively---fixes $\alpha =2\beta $, reducing this equation to $(1+2\beta
)\Gamma \ +6[\Gamma \frac{\partial ^{2}\Gamma }{\partial \zeta ^{2}}+(\frac{%
\partial \Gamma }{\partial \zeta })^{2}]$ $+(1-\beta )$ $\zeta \frac{%
\partial \Gamma }{\partial \zeta }$ $=0$. Substituting $H^{\ast
}(T;X,X_{0})=T^{-1/3}\eta ^{2\beta }\Gamma (\xi /\eta ^{\beta })$ (with $\xi
=X/T^{1/3}$ and $\eta =X_{0}/T^{1/3}$) into the mass conservation law yields 
$1$ $=$ $\int_{-\infty }^{\infty }H^{\ast }(T;X,X_{0})\,dX$ $=\left( \frac{%
X_{0}}{T^{1/3}}\right) ^{3\beta }$ $\int_{-\infty }^{\infty }\Gamma (\zeta
)\,d\zeta $, which requires $\beta =0$ so that the total mass remains
conserved in time (and independent of $X_{0}$). Setting $\beta =0$ reduces
the differential equation above to the same form as in the single-variable
analysis and recovers Eqs \ref{eq5} and \ref{eq6}.

\subsection{Note on the modified DDD model\label{A-MDD}}

We first discuss the necessity and sufficiency of the scale-covariance
identity in Eq \ref{eq3b}. Consider the fixed-point solution $H^{\ast }$ in
Eq \ref{eq3a}. For $H^{\ast }$ to remain self-similar as time evolves, any
time rescaling $T\rightarrow \sigma T$ ($\sigma >0$) must reproduce the
spatial profile up to multiplicative rescalings of the spatial coordinate
and amplitude: $H^{\ast }(\sigma T,\phi (\sigma )X,X_{0})$ \ $=$ $\psi
(\sigma )H^{\ast }(T,X,X_{0})$, where $\phi (\sigma ),\psi (\sigma )$ ($>0$)
are continuous scale factors. Applying two successive rescalings $\sigma
_{1} $ and $\sigma _{2}$ must coincide with a single rescaling by $\sigma
_{1}\sigma _{2}$, which enforces $\phi (\sigma _{1}\sigma _{2})=\phi (\sigma
_{1})\phi (\sigma _{2})$ and $\psi (\sigma _{1}\sigma _{2})=\psi (\sigma
_{1})\psi (\sigma _{2})$, which result in $\phi (\sigma )=\sigma ^{\mu }$
and $\psi (\sigma )=\sigma ^{\nu }$. Substituting Eq \ref{eq3a} into $%
H^{\ast }(\sigma T,\sigma ^{\mu }X,X_{0})$ $\ =$ $\sigma ^{\nu }H^{\ast
}(T,X,X_{0})$ and then dividing both sides by $(\sigma T)^{-A/B}$ yields the
scale-covariance identity in Eq \ref{eq3b}, where $\lambda =\sigma ^{-1/B}$, 
$\beta =1-\mu B$, and $\alpha =-A-\nu B$. Conversely, setting $\lambda
=1/\eta $ and defining $\Gamma (\zeta )=f(\zeta ,1)$ with $\xi /\eta ^{\beta
}$ immediately gives Eq \ref{eq23}, so that $H^{\ast
}(T,X,X_{0})=T^{-(A+\alpha )/B}X_{0}^{\alpha }\Gamma (X/(T^{(1-\beta
)/B}X_{0}^{\beta }))$ collapses onto a single invariant master profile $%
\Gamma (\zeta )$ at all times $T$ and initial scales $X_{0}$. (Note that in
the fatigue problem of Sec \ref{sec:fatigue}, where the evolution variable $%
T $ does not involve the arbitrary reference length $l_{\mathrm{ref}}$,
unit-choice invariance directly enforces the isotropic special case of $%
\beta =1$ and $\alpha =1$ without requiring an additional time-evolution
self-similarity condition.)

For the extra term $\left( \partial _{X}H\right) ^{2}$ with $N_{11}=2$, the
scaling exponent is calculated as $M=A+B-2(A+1)=B-A-2$; this extra term is
scale-invariant. The scale invariance of these terms gives a condition $%
B=A+2 $. To derive Eq \ref{eq10} from Eqs \ref{eq11} and \ref{eq23}, we use
the chain-rule relations $\frac{\partial f}{\partial \xi }=\eta ^{\beta }%
\frac{d\Gamma }{d\zeta }$ and $\partial _{\xi }(f\frac{\partial f}{\partial
\xi })$ $=$ $\eta ^{2\beta }[\Gamma \frac{d\Gamma }{d\xi }+(\frac{d\Gamma }{%
d\xi })^{2}]$, and $\xi \frac{\partial f}{\partial \xi }+\eta \frac{\partial
f}{\partial \eta }=\eta ^{2\beta }[(1-\beta )\zeta \frac{d\Gamma }{d\zeta }%
+\alpha \Gamma ]$ (note $\alpha =2\beta $). Substituting these into Eq \ref%
{eq11} cancels the common prefactor $\eta ^{2\beta }$ across all terms,
yielding $(A+2\beta )\Gamma $ $+2B$ $[\Gamma \frac{d^{2}\Gamma }{d\zeta ^{2}}
$ $+(1-c)(\frac{d\Gamma }{d\zeta })^{2}]$ $+(1-\beta )\xi \frac{d\Gamma }{%
d\xi }$ $=0$; dividing through by $(1-\beta )$ and setting $B=(1-\beta )b$
then yields Eq \ref{eq10}.

Rescaling $\zeta $ so as to normalize the front position to $\zeta =1$(i.e.,
setting $X_{f}=1$), the three boundary conditions to be satisfied by the
second-order differential equation in Eq \ref{eq10} are given as follows: $%
d\Gamma /d\zeta =0$ at $\zeta =0$, $\Gamma (\zeta )=0$ at the front $\zeta
=1 $, and a nontrivial condition $2b(1-c)d\Gamma /d\zeta =-1$ at $\zeta =1$
(although the original argument in the text \cite{barenblatt2003scaling} is
highly involved, the last condition may simply be justified by setting $%
\zeta =1$ and $\Gamma =0$ in Eq \ref{eq10}).

Equation \ref{eq8} can be confirmed by substituting $\Gamma (\zeta
)=C(X_{f}^{2}-\zeta ^{2})$ into Eq \ref{eq10} to have $2bC^{2}$ $%
[-2(X_{f}^{2}-\zeta ^{2})$ $+(1-c)(2\zeta )^{2}]$ $-2C\zeta ^{2}$ $%
+Cp(X_{f}^{2}-\zeta ^{2})$ $=0$, from which we obtain $-4bC+p=0$ and $%
4bC(3-2c)-2-p=0$. This shows the existence of the solution of the above form
for Eq \ref{eq8}.

In the present case, the mass is not conserved: $\int_{-\infty }^{\infty
}H(T,X)dX=2\frac{X_{0}^{2\beta }}{T^{\frac{A+2\beta }{B}}}\int_{0}^{T^{\frac{%
1-\beta }{B}}X_{0}^{\beta }}dX\Gamma (\frac{X}{T^{\frac{1-\beta }{B}%
}X_{0}^{\beta }})\sim 2\frac{X_{0}^{3\beta }}{T^{\frac{A+3\beta -1}{B}}}$.
In a dimensional form $\int_{-\infty }^{\infty }h(t,x)dx\sim 2\frac{%
m_{0}^{1-\beta }l^{3\beta }}{(\kappa t)^{\beta }}$, which shows that $x_{f}^{%
\frac{3\beta }{1-\beta }}\int_{-\infty }^{\infty }h(t,x)dx$ is a conserved
quantity.

\subsection{Note on Barenblatt's nonlinear diffusion\label{A-BD}}

Barenblatt's nonlinear diffusion equation models filtration in an
elastoplastic porous medium, where the effective diffusivity jumps
discontinuously depending on whether the medium is unloading ($\partial
_{t}h<0$) or loading ($\partial _{t}h>0$)~\cite{Barenblatt, Goldenfeld}: 
\begin{equation}
\partial _{t}h(t,x)=\kappa _{\mathrm{eff}}\,\partial _{x}^{2}h(t,x),\qquad
\kappa _{\mathrm{eff}}=%
\begin{cases}
\kappa (1+\epsilon ) & \text{for }\partial _{t}h<0\quad (|x|<x_{\ast }(t)),
\\ 
\kappa & \text{for }\partial _{t}h>0\quad (|x|>x_{\ast }(t)),%
\end{cases}%
\end{equation}%
with $\epsilon >-1$ ($\epsilon \neq 0$) and $\pm x_{\ast }(t)$ denoting the
turning points where $\partial _{t}h=0$. Introduction of dimensionless
variables, $h_{0}=m_{0}/l_{0}$, $X=x/l_{0}$, $X_{0}=l/l_{0}$, and $T=\kappa
t/l_{0}^{2}$, renders this governing equation into 
\begin{equation}
\partial _{T}H=D\,\partial _{X}^{2}H,\qquad D=%
\begin{cases}
1+\epsilon & \text{for }\partial _{T}H<0\quad (|X|<X_{\ast }(T)), \\ 
1 & \text{for }\partial _{T}H>0\quad (|X|>X_{\ast }(T)),%
\end{cases}%
\end{equation}%
where $X_{\ast }(T)=x_{\ast }(t)/l_{0}$. Because $D$ depends solely on the
sign of $\partial _{T}H$, the operator $D\,\partial _{X}^{2}H$ remains
invariant under the rescaling in Eqs \ref{eq14}-\ref{eq16} with $M=B-2=0$
(fixing $B=2$), whereas the step discontinuity $\epsilon \neq 0$ breaks mass
conservation and gives rise to the anomalous dimension $\alpha =\alpha
(\epsilon )$.

Starting from the identification $H(T;\mathbf{X})=H(T;X,X_{0})$, the
stationary RG flow equation is given not by $0=Af(\xi )$ $+\xi \frac{%
\partial f(\xi )}{\partial \xi }$ $+BD\frac{\partial ^{2}f(\xi )}{\partial
\xi ^{2}}$, but by $0=$ $Af(\xi ,\eta )$ $+\xi \frac{\partial f(\xi ,\eta )}{%
\partial \xi }$ $+\eta \frac{\partial f(\xi ,\eta )}{\partial \eta }$ $+BD%
\frac{\partial ^{2}f(\xi ,\eta )}{\partial \xi ^{2}}$. Applying the $\lambda 
$-rescaling invariance method introduced in Sec \ref{sec:mddd} yields Eq \ref%
{eq23} (here with $\alpha $ replaced with $\widetilde{\alpha }$, for
convenience) and fixes $\beta =0$ (corresponding to a memory-retaining I
fixed point), because invariance of the stationary RG flow equation---whose
linear and diffusive terms scale as $\lambda ^{\widetilde{\alpha }}$ and $%
\lambda ^{\widetilde{\alpha }-2\beta }$, respectively---requires $\widetilde{%
\alpha }=\widetilde{\alpha }-2\beta $, resulting in $\beta =0$. The method
also reduces the equation to $0=$ $(A+2\alpha )\Gamma $ $+\zeta \frac{%
\partial \Gamma (\zeta )}{\partial \zeta }$ $+BD\frac{\partial ^{2}\Gamma
(\zeta )}{\partial \zeta ^{2}}$, where we have set $\widetilde{\alpha }%
=2\alpha $ to match the standard convention for Barenblatt's anomalous
dimension $\alpha $ \cite{Goldenfeld}.

Here, $B=2$ is fixed by the scale invariance of the diffusion equation ($%
M=B-2=0$). Furthermore, under the standard nondimensionalization $%
h_{0}=m_{0}/l_{0}$, $X=x/l_{0}$, $X_{0}=l/l_{0}$, and $T=\kappa t/l_{0}^{2}$%
, the dimensional solution $h^{\ast }(t,x,l)=h_{0}T^{-A/2}\eta ^{2\alpha
}\Gamma (\xi )$ carries a prefactor proportional to $l_{0}^{A-1}$; requiring
invariance under the choice of the arbitrary reference length $l_{0}$
uniquely pins down $A=1$, while the eigenvalue problem fixes the combination 
$A+2\alpha $ (and hence $\alpha $). Setting $A=1$ and $B=2$ in this equation
yields $0=$ $(1+2\alpha )\Gamma $ $+\zeta \frac{\partial \Gamma (\zeta )}{%
\partial \zeta }$ $+2D\frac{\partial ^{2}\Gamma (\zeta )}{\partial \zeta ^{2}%
}$, which coincides with the stationary RG flow equation obtained from the
single-variable identification $\mathbf{H}(T;\mathbf{X})=H(T,X)$ (where the
anomalous dimension is instead absorbed into $A=1+2\alpha $ \cite{Goldenfeld}%
). The anomalous dimension $\alpha $ was first obtained numerically by
Barenblatt \cite{Barenblatt} and later calculated analytically, though
perturbatively, by Goldenfeld and co-workers using field-theoretic RG
methods \cite{chen1994renormalization}.

\subsection{Derivation of the Three-Scale RG Fixed Point in Fatigue Crack
Growth}

\label{app:fatigue}

In this appendix, we provide the detailed RG scaling analysis of
Barenblatt's fatigue-crack growth problem leading to Eqs.~%
\eqref{fatigue_fp_eq}--\eqref{fatigue_Hstar}.

\subsubsection{Dimensional Formulation and Relevance of Power-Law Terms}

In a fatigue test on a pre-notched specimen of characteristic size $h$, the
stress-intensity factor is $N = c\sigma\sqrt{l}$ and its cycle amplitude is $%
\Delta N = N_{\max} - N_{\min} = c\Delta\sigma\sqrt{l}$. Dimensional
analysis of the cycle-averaged crack growth rate $dl/dn$ in terms of $\Delta
N $, $R = \sigma_{\max}/\sigma_{\min}$, yield stress $\sigma_{Y}$, fracture
toughness $K_{Ic}$, and specimen size $h$ gives \cite{barenblatt2014flow} 
\begin{equation}
\frac{dl}{dn} = \left(\frac{\Delta N}{\sigma_{Y}}\right)^{2} \Phi\!\left(%
\frac{\Delta N}{K_{Ic}}, R, Z\right), \qquad Z = \frac{\sigma_{Y}\sqrt{h}}{%
K_{Ic}} = \sqrt{\frac{h}{l_{c}}},  \label{app_fatigue_dim}
\end{equation}
where $l_{c} \equiv (K_{Ic}/\sigma_{Y})^{2}$ and $\Phi$ is an unspecified
dimensionless function. Substituting $\Delta N = c\Delta\sigma\sqrt{l}$ and $%
\varepsilon \equiv c\Delta\sigma/\sigma_{Y}$ yields Eq.~\eqref{fatigue_ode},
which in terms of $H(T; X_{0}, X_{1}, X_{2}) = l(n)/l_{\mathrm{ref}}$ and $T
= (n_{f} - n)/t_{\mathrm{ref}}$ ($1/t_{\mathrm{ref}} = \varepsilon^{2}$)
takes the form of Eq.~\eqref{fatigue_rg_pde}.

From the predictive RG viewpoint \cite{Okumura2026nonlinear}, we examine Eq.~%
\eqref{fatigue_rg_pde} under the Case~I ($0<L<1$) rescaling $(T,\mathbf{X}%
)\rightarrow (L^{B}T,L\mathbf{X})$ in Eq.~\eqref{eq16}, where $H(L^{B}T,L%
\mathbf{X})=L^{-A}H^{L}(T,\mathbf{X})$. Under this transformation, $Z=\sqrt{%
X_{2}/X_{1}}$ is invariant, whereas $u=\varepsilon \sqrt{H/X_{1}}$ scales as 
$u\rightarrow L^{-(A+1)/2}u$. Expanding $\Phi (u,R,Z)=\sum_{k}\Phi
_{k}(R,Z)u^{k}$, the $k$-th term $-H\Phi _{k}(R,Z)u^{k}=-\Phi
_{k}(R,Z)\varepsilon ^{k}H^{1+k/2}X_{1}^{-k/2}$ acquires the scaling factor $%
L^{M_{k}}$ with exponent (cf.~Eq.~\eqref{M1}) $M_{k}$ $=A$ $+B$ $-A\!\left(
1+\frac{k}{2}\right) $ $-\frac{k}{2}$ $=B$ $-\frac{k}{2}(A+1)$. Scale
invariance of a leading power $k=\alpha (R,Z)>0$ ($m=2+\alpha >2$) fixes $B=$
$\frac{\alpha (R,Z)}{2}$ $(A+1)$ and gives $M_{k}=\frac{1}{2}(A+1)[\alpha
(R,Z)-k]$ for all other powers. Since $B>0$ and $\alpha (R,Z)>0$, we have $%
A+1>0$. In Case~I ($0<L<1$), lower-order terms with $k<\alpha (R,Z)$ have $%
M_{k}>0$ and are \textit{irrelevant}, vanishing under repeated RG
transformations ($L^{nM_{k}}\rightarrow 0$ as $n\rightarrow \infty $),
whereas higher-order terms with $k>\alpha (R,Z)$ have $M_{k}<0$ and are 
\textit{relevant}, so they must be absent from the effective fixed-point
generator in the intermediate regime $\Delta N\ll K_{Ic}$. Eliminating the
irrelevant terms leaves the effective scale-invariant equation 
\begin{equation}
\partial _{T}H=-\varepsilon ^{\alpha (R,Z)}\Phi _{1}(R,Z)H^{1+\alpha
(R,Z)/2}X_{1}^{-\alpha (R,Z)/2}.  \label{app_fatigue_asymp}
\end{equation}%
The RG framework predicts for the field $H$ governed by this equation the RG
fixed-point of the form $H^{\ast }(T;X_{0},X_{1},X_{2})=T^{-A/B}f(\eta
_{0},\eta _{1},\eta _{2})$ if the solution to Eq \ref{fatigue_fp_eq} for $%
f(\eta _{0},\eta _{1},\eta _{2})$ exists.

\subsubsection{Scale Selection from the Stationary RG Flow Equation}

In order to solve Eq \ref{fatigue_fp_eq}, we apply the same $\lambda $%
-rescaling invariance method as in Sec \ref{sec:mddd}, utilizing that this
equation must be invariant under a change of the arbitrary reference length $%
l_{\mathrm{ref}}\rightarrow l_{\mathrm{ref}}/\lambda $ with $\lambda >0$.
Because $T$, $R$, $Z=\sqrt{\eta _{2}/\eta _{1}}$, and $\varepsilon $ do not
contain $l_{\mathrm{ref}}$, they remain invariant under this rescaling,
whereas $\eta _{i}\rightarrow \lambda \eta _{i}$ ($i=0,1,2$) and $%
f\rightarrow \lambda ^{\gamma }f$, i.e., $f(\lambda \eta _{0},\lambda \eta
_{1},\lambda \eta _{2})=\lambda ^{\gamma }f(\eta _{0},\eta _{1},\eta _{2})$,
with $\gamma $ to be determined. Under this transformation, the first and
third terms of Eq \ref{fatigue_fp_eq}, which are linear in $f$, are
multiplied by $\lambda ^{\gamma }$, while the second nonlinear term is
multiplied by $\lambda ^{\gamma (1+\alpha /2)-\alpha /2}$ owing to the
factor $f(\sqrt{f/\eta _{1}})^{\alpha }$. Scale invariance of Eq \ref%
{fatigue_fp_eq} therefore requires $\gamma =\gamma (1+\alpha /2)-\alpha /2$,
which uniquely fixes $\gamma =1$ (consistent with the $l_{\mathrm{ref}}$%
-independence of the dimensional crack length $l^{\ast }(n)=l_{\mathrm{ref}%
}T^{-A/B}f(\eta _{0},\eta _{1},\eta _{2})$), giving $f(\lambda \eta
_{0},\lambda \eta _{1},\lambda \eta _{2})=\lambda f(\eta _{0},\eta _{1},\eta
_{2})$.

Setting $\lambda =1/\eta _{1}$, we obtain $f(\eta _{0}/\eta _{1},1,\eta
_{2}/\eta _{1})=f(\eta _{0},\eta _{1},\eta _{2})/\eta _{1}$, i.e., $f(\eta
_{0},\eta _{1},\eta _{2})=\eta _{1}\Gamma (\eta _{0}/\eta _{1},\eta
_{2}/\eta _{1})$. From this or by taking the derivative with respect to $%
\lambda $ of both sides of $f(\lambda \eta _{0},\lambda \eta _{1},\lambda
\eta _{2})=\lambda f(\eta _{0},\eta _{1},\eta _{2})$ and then setting $%
\lambda =1$, we obtain $\sum_{i=0}^{2}\eta _{i}\frac{\partial f}{\partial
\eta _{i}}=f$. This equation changes Eq \ref{fatigue_fp_eq} into a simple
equation for $f$, which readily gives $\Gamma =\left( \frac{2}{\alpha
(R,Z)\,\Phi _{1}(R,Z)\,\varepsilon ^{\alpha (R,Z)}\,}\right) ^{2/\alpha
(R,Z)}$ by using $B=\frac{1}{2}\alpha (R,Z)(A+1)$. From this we see $\Gamma
(\eta _{0}/\eta _{1},\eta _{2}/\eta _{1})$ does not depend on $\eta
_{0}/\eta _{1}$ but depends on $Z=\sqrt{\eta _{2}/\eta _{1}}$: 
\begin{equation}
f(\eta _{0},\eta _{1},\eta _{2})=\eta _{1}\Gamma (\eta _{2}/\eta _{1}),
\label{app_fatigue_ansatz}
\end{equation}%
which results in Eqs \ref{eqA} and \ref{fatigue_Hstar}.

Based on Eq \ref{fatigue_Hstar}, differentiating $l^{\ast }(n)=l_{\mathrm{ref%
}}H^{\ast }$ with respect to $n$ via $dl^{\ast }/dn=-(l_{\mathrm{ref}}/t_{%
\mathrm{ref}})\partial _{T}H^{\ast }$ and eliminating $T$ in favor of $%
l^{\ast }$ recovers 
\begin{equation}
\frac{dl^{\ast }}{dn}=(\varepsilon \sqrt{l^{\ast }})^{2+\alpha (R,Z)}\Phi
_{1}(R,Z)\,\,l_{c}^{-\alpha (R,Z)/2}=\frac{\Phi _{1}(R,Z)}{\sigma
_{Y}^{2}K_{Ic}^{\alpha (R,Z)}}(\Delta N)^{2+\alpha (R,Z)},
\end{equation}%
which reproduces Barenblatt's Paris scaling law \cite{barenblatt2014flow}.
Note that the second equality is justified by $\Delta N=c\Delta \sigma \sqrt{%
l^{\ast }}=\sigma _{Y}\varepsilon \sqrt{l^{\ast }}$.

\end{document}